\documentclass[aps,amsmath,twocolumn,amssymb,floatfixng,showpacs,superscriptaddress,footinbib]{revtex4-1}

\usepackage[utf8]{inputenc}
\usepackage{amsmath}
\usepackage{amsfonts}
\usepackage{xcolor}
\usepackage{natbib}
\usepackage{graphicx}
\usepackage{physics}
\usepackage{bm}
\usepackage{cleveref}
\usepackage{listings}
\usepackage{cases}
\usepackage{tikz}

\newcommand\bea{\begin{eqnarray}}
\newcommand\eea{\end{eqnarray}}
\newcommand\beq{\begin{equation}}  
\newcommand\eeq{\end{equation}}

\usepackage[normalem]{ulem}
\usepackage{sidecap,tikz}
\DeclareRobustCommand{\orcidicon}{\hspace{-1.0mm}
	\begin{tikzpicture}
	\draw[lime, fill=lime] (0.0,0.0) 
	circle [radius=0.15] 
	node[white] {{\fontfamily{qag}\selectfont \tiny \,ID}};
	\draw[white, fill=white] (-0.0525,0.095) 
	circle [radius=0.007];
	\end{tikzpicture}
	\hspace{-3.0mm}
}
\foreach \x in {A, ..., Z}{\expandafter\xdef\csname orcid\x\endcsname{\noexpand\href{https://orcid.org/\csname orcidauthor\x\endcsname}
		{\noexpand\orcidicon}}
}

\begin{document}
\title{Fano resonances for tilted linear and quadratic band touching dispersions in a harmonically driven potential well}
\author{Anton Gregefalk}\email{angr1126@gmail.com}\affiliation{Department of Physics and Astronomy, Uppsala University, Box 516, 75120 Uppsala, Sweden}
\author{Annica M. Black-Schaffer}
\affiliation{Department of Physics and Astronomy, Uppsala University, Box 516, 75120 Uppsala, Sweden}
\author{Tanay Nag\orcidB{}}
\email{tanay.nag@physics.uu.se}\affiliation{Department of Physics and Astronomy, Uppsala University, Box 516, 75120 Uppsala, Sweden}
\affiliation{Department of Physics, BITS Pilani-Hyderabad Campus, Telangana 500078, India}

\begin{abstract}
Considering models with tilted linear and quadratic band touching dispersions, we analyze the effect of the transverse linear tilt on the transmission spectra through a harmonically driven potential well oriented longitudinally. Employing the Floquet scattering matrix formalism, we find Fano resonances as an outcome of matching between the Floquet sidebands and quasi-bound states, where the tilt renormalizes their energies and wave vectors. We find that the Fano resonance energy decreases (increases) for linear (quadratic) band touchings as the magnitude of the transverse momentum increases, indicating a distinct signature of the underlying band dispersion in the transmission profile. The sign of the product of the transverse momentum and the tilt also determines the relative shift in the Fano resonance energy with respect to the untilted case for both band dispersions, suggesting a possible tunability of the Fano resonance for tilted systems. Importantly, the tilt strength can also be directly determined by measuring the Fano resonance energy as function of the transverse momenta direction.
We furthermore study the shot noise spectra and their differential property where we find an inflection region and undulation, respectively, around the Fano resonance energy. 
Interestingly, differential shot noise and transmission spectra both qualitatively behave in a similar fashion and might thus serve as important observables for future experiments on driven solid-state systems. 

\end{abstract}

\maketitle 

%%%%%%%%%%%%%%%%%%%%%%%%%%%%%%%%%%%%%%%%%%%%%%%%%%%%%%%%%%%%%%%%%%%%
\section{Introduction}
\label{sec1}
%%%%%%%%%%%%%%%%%%%%%%%%%%%%%%%%%%%%%%%%%%%%%%%%%%%%%%%%%%%%%%%%%%%%

%%%%%%%%%%%%%%%%%%%  Floquet and Floquet scattering %%%%%%%%%%%%%%%%%%%%%%
Periodically driven isolated systems have emerged as an exciting field of research in recent years due to their intriguing properties as compared to their equilibrium counterparts \cite{shirley65,dunlap86,grifoni98}. A few examples  of such nonequilibrium phenomena are dynamical localization \cite{kayanuma08,nag14,nag15,Tamang21}, many-body localization \cite{d13many,d14long,ponte15periodically,ponte15,lazarides15fate,zhang16}, quantum phase transitions \cite{eckardt05superfluid,zenesini09}, Floquet topological insulators \cite{oka09photovoltaic,kitagawa11transport,lindner11floquet,rudner13anomalous,vega19,seshadri19,nag19,nag20a,nag20b}, Floquet topological superconductors \cite{ghosh21a,ghosh21b,Mondal23},
Floquet time crystals \cite{else16floquet,khemani16phase,zhang17observation,yao17discrete}, and higher harmonic generation \cite{faisal97,nag17,ikeda18,neufeld2019floquet}. For many of these systems a fundamental underlying question is how the electrons, expressed as wave packets, interact with a time-dependent 
quantum well. In this context, a Fano resonance, caused by destructive interference between a discrete bound state from the well interacting with a continuum of propagating modes,  becomes an important physical phenomenon and they are already widely observed in atomic spectrum, light propagation, quantum transport, matter-wave scattering in ultracold atom systems, etc \cite{Miroshnichenko10,Fano61,Tekman93}. For example, current measurements often reveal interesting features  that are mediated by these Fano resonances. 
On the other hand, current fluctuations, originating from the quantization of charge carriers, is captured by the shot noise \cite{blanter2000shot,Lefloch03,Moskalets04,Saminadayar97}. In the case of a time-dependent quantum well, apart from Fano resonances in the transmission spectrum, there also exist extensive investigations on pumped shot noise and Wigner-Smith delay times \cite{Wenjun99,Agapi02,zhu2015fano,zhu2017fano,biswas2017electron,Yonatan21,bera2021floquet,Longhi15}. Overall, the scattering properties of time-varying potential barriers or wells or even laser-driven potentials \cite{Agapi02} have been found to lead to various interesting outcomes, such as  photon-assisted tunneling \cite{platero2004photon,pimpale1991tunnelling}, and quantum pumping \cite{zhu2009quantum,Moskalets04}. 

%%%%%%%%%%%%%%%%%%%  linear and non-linear dispersion, band topology, tilt and others %%%%%%%%%%%%%%%%%%%%%%%%%%%%%%%%%
The band dispersion of a material, captured by tight-binding models as well as first principle investigations, 
is a central concept in condensed matter physics as it determines many of the physical properties.  The advent of graphene, its derivatives, and surface states of topological insulators have allowed us to investigate linear band touching (LBT) dispersions in two-dimensional (2D) systems \cite{Louvet15,Illes15,Mukherjee15,biswas2016magnetotransport,Neto09,geim2007rise,Hasan10,Xiao-Liang11}. For three-dimensional (3D) systems, the Weyl and Dirac semimetals are shown to exhibit LBTs as well \cite{Armitage18}. Non-linearity in the band dispersion can be obtained for multilayer graphene. For example, bilayer and trilayer graphene harbor quadratic band touching (QBT) and cubic band dispersions, respectively \cite{geisenhof2021quantum,velasco2012transport,lee2014competition}. Interestingly,  QBTs have also been found in 2D \cite{Kai09,tsai2015interaction,Das21} and 3D \cite{kondo2015quadratic,Moon13} systems. 
It is also possible to engineer various dispersions through different lattices, such as the Lieb lattice, kagome lattice, dice or $T_3$ lattice, and $K_4$ crystal used to investigate the interplay between flat, linear, or non-linear bands. 
Overall band engineering has recently received a lot of attention due to the experimental advancement in creating optical lattices based on cold atoms \cite{Bloch08}, metamaterials constructed out of photonic and phononic crystals \cite{li2019topology}, and even solid compounds through van der Waals stacking and heterostructures \cite{Nakayama12,zhang2016efficient}. 

\begin{figure}
    \centering
    \includegraphics[width = \linewidth]{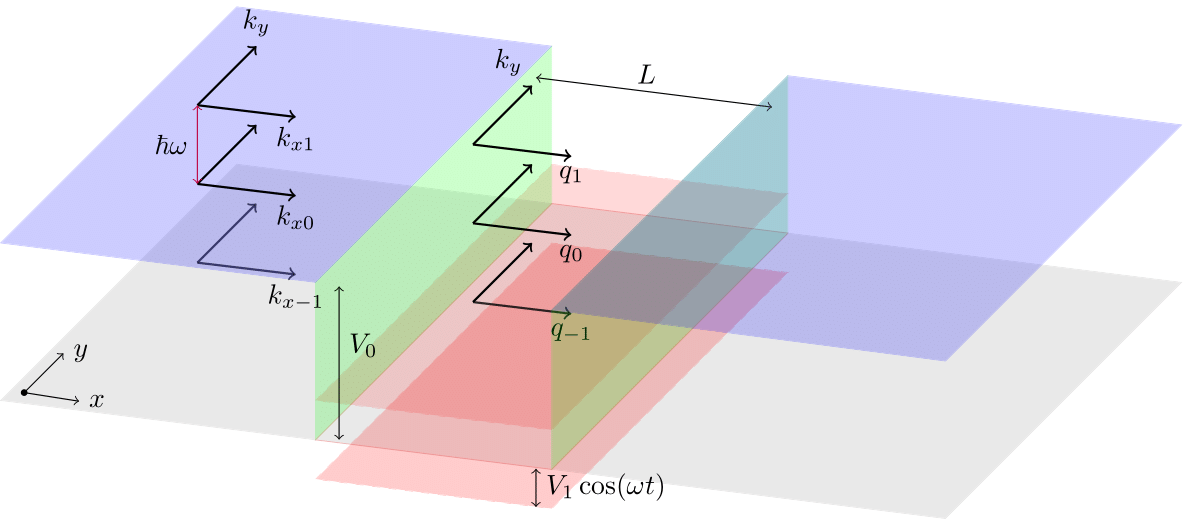}
    \caption{Schematic potential well in the $xy$-plane. Electrons are transported in the $x$-direction along which a potential well of depth \(V(t) = V_0 + V_1\cos(\omega t)\) and width \(L\) is oriented longitudinally.  The potential well is infinitely extended in the transverse \(y\)-direction, which is also the tilt direction as described in Eqs.~(\ref{eq:Hamiltonian-graphene}) and (\ref{eq:Hamiltonian-QBT}). \textcolor{black}{Here $k_{xn}$ ($q_{xn}$) represents the wave vectors associated with the free (potential) region for the $n=0, \pm 1$ Floquet side bands.}}
    \label{fig:setup}
\end{figure}

Interestingly, a leaning of the energy dispersion along a particular momentum direction, but without opening up a gap, referred to as a tilt, not only changes the dispersion but can also have far reaching consequences on properties, such as
generating an anisotropic spin texture, Fermi surface topology, Hall currents, etc \cite{Zhang13b,soluyanov2015type,Ma19}. Examples of tilted LBT materials include an in-plane magnetic field on the surface of the topological insulator introducing a tilt in the linear dispersion \cite{Zhang12,Zhang13b,nag2022third}, while 
tilt can easily be intrinsically present in Dirac and Weyl semimetals with linear as well non-linear band dispersions \cite{Nag20,Nag22,Sadhukhan23,Das21}.
Periodically driven quantum wells for systems with linear and quadratic band touchings have already been extensively investigated in the absence of tilt \cite{zhu2015fano,bera2021floquet,bera2023floquet}, but the effects of the often present tilt have so far not been studied.
Therefore, we are here interested in analyzing  the effect of  tilt in both LBT and QBT dispersive materials on the transmission spectra. Keeping in mind a Hall setup, we consider a perpendicular arrangement between the quantum well i.e., transmission direction and the tilt direction, see Fig. \ref{fig:setup} for a schematic setup. \textcolor{black}{
To be precise, in this work we address the following questions:} How do the quasi-bound states, associated with the potential well in the longitudinal direction, interact with the Floquet sidebands produced due to the periodic driving in the presence of a transverse tilt in the dispersion of LBT and QBT materials? And how do positive and negative tilt modify the Fano resonance and the shot noise spectra with respect to the untilted case? 

%%%%%%%%%%%%%%%%%%%%%%%%%%%%%%%%%%%%%%%%%%%%%%%%%%%%%%%%%%%%
%%%%%%%%%%%%%%%%%%%%%%%%%%%%%%%%%%%%%%%%%%%%%%%%%%%%%%%%%%%%%%

%%%%%%%%%%%%%%%%%%%%%%%  brief description of our work  %%%%%%%%%%%%%%%%%%%%%%%%%%%%%%%%%%%
In particular, in this work we investigate the transmission  spectra for a harmonically driven potential well considering tilted LBT and QBT systems (see Figs. \ref{fig:graphene-band-tilt} and \ref{fig:qbt-band-tilt}). We find that the Fano resonance energy decreases and increases  for LBT and QBT, respectively, when increasing the magnitude of the transverse tilt momenta (see Figs. \ref{fig:graphene-transmission-posky},  \ref{fig:graphene-transmission-negky}, \ref{fig:qbt-transmission-posky}, and \ref{fig:qbt-transmission-negky}). 
Interestingly, positive (negative) tilt always shifts the Fano resonance energy upward (downward) for positive transverse momenta, while the findings are reversed for negative transverse momenta. These results hence indicate that the product of the tilt and transverse momenta is the important factor in engineering the Fano resonance energy.  Importantly, by just identifying the Fano resonance peaks for positive and negative transverse motion we can even determine the strength of the tilt, without any prior knowledge of the tilt. Furthermore, the qualitative differences between the Fano resonances also clearly distinguishes between the LBT and QBT dispersions.
We furthermore study the shot noise and differential shot noise spectra where we find an inflection region and clear undulation, respectively, marking the emergence of the Fano resonance (see Figs. \ref{fig:graphene-shotnoise-posky},  \ref{fig:graphene-shotnoise-negky}, \ref{fig:qbt-shotnoise-posky}, and \ref{fig:qbt-shotnoise-negky}). Notably, the  differential shot noise and transmission spectra behave in an similar manner. This additionally suggests that the overlap between dominant Floquet sidebands and renormalized energies of the quasi-bound well states is essential to observe Fano resonances.

%%%%%%%%%%%%%%%%%%%% construction %%%%%%%%%%%%%%%%%%%%%

The remaining of this article is organized as follows. Sec.~\ref{sec2} 
discusses the models of LBT and QBT dispersions and the methods for solving for the transmission spectra and and shot noise. In Sec.~\ref{sec3} we report our numerical results where the effects of tilt and transverse momentum are extensively studied in both transmission and shot noise spectra for LBT and QBT systems.  In Sec.~\ref{sec4} we analyze the results, especially in terms of the underlying band dispersion and provide plausible arguments to understand the numerical results. \textcolor{black}{ We discuss the limitations of our work with regard to realistic mesoscopic devices in Sec.~\ref{sec_new}.}
At the end, in  Sec.~\ref{sec5}, we conclude with possible future scopes.

%%%%%%%%%%%%%%%%%%%%%%%%%%%%%%%%%%%%%%%%%%%%%%%%%%%%%%%%%%%%%%%%%%%%
\section{Model and method}
\label{sec2}
%%%%%%%%%%%%%%%%%%%%%%%%%%%%%%%%%%%%%%%%%%%%%%%%%%%%%%%%%%%%%%%%%%%%

In this work we consider both LBT and QBT dispersions. A general form of the Hamiltonian for both of these dispersions is \(  H=\bm{n}\cdot \bm{\sigma} + n_0 (k_i) \sigma_0 \) with \(\bm{n}=(n_x(\bm{k}), n_y(\bm{k}), n_z(\bm{k}) ) \),  \( \bm{\sigma} =(\sigma_x, \sigma_y, \sigma_z) \) representing the Pauli matrices, and \(n_0(k_i)\) encoding the tilt  that cants the spectrum along the \( k_i\)-direction, with $\sigma_0$ denoting the identity matrix. We here limit ourselves to 2D systems, such that $\bm{k}=(k_x,k_y)$. The specific form of $\bm{n}$ is given in Secs. \ref{sec2_1} and \ref{sec2_2}, respectively, LBT and QBT.  
The potential landscape for the periodically driven quantum well can be described as: 
\[
    V(r_j,t) = \begin{cases} 
    V_0+V_1\cos(\omega t)\quad \text{if } \quad \abs{r_j} \leq L/2, \forall {r_i}\\
    0\quad \quad \text{elsewhere},
    \end{cases}
\]
where \(V_0<0\) is amplitude of the the static potential, \(V_1\) the amplitude of the driving  and \(\omega\) its frequency. We assume that $|V_0|\gg |V_1|$ to make sure a quantum well is always present in the central region.
The potential is confined along the transverse  \( j \)-th direction 
\(\abs{r_j} \leq L/2\) with respect to the tilt in \(i\)-th direction, where \( i\ne j\). Note that the potential has an infinite extension along the tilt direction. See Fig. \ref{fig:setup} for a schematic representation of the potential well. We further consider \(n_0(k_i)= \hbar \tau_y k_y\) and \( V(r_j,t)= V(x,t) \) for our analysis, where $\tau_y$ accounts for the tilt strength.

Due to translation invariance along the $y$-direction, we can generally assume $
\Psi(x,y) = e^{ik_yy}\psi(x,t) $
for the \textcolor{black}{single-particle} solutions of the time-dependent Schr\"odinger equation. Further, due to the harmonic modulation of the potential well with amplitude $V_1$ and frequency $\omega$, we can solve the problem within the Floquet formalism. As a result, we work with Floquet sidebands of energy $E_l= E_F + l\hbar \omega $, where $l \in \mathbb{Z}$ labels the the Floquet sidebands and $E_F$ denotes the incident energy of the wave packet. Given the  specific choice of Hamiltonian, we formulate $\psi(x,t)$ as described below in Eqs. (\ref{eq:tiltgraphenewf}) and (\ref{eq:tiltwf}).
\\

Without loss of generality, we can consider the incoming wave as $\bm{A}^i_m, \bm{B}^i_m$ for the $m$-th sideband and the outgoing wave as $\bm{A}^o_n, \bm{B}^o_n$ for the $n$-th sideband, with respect to the scattering potential $V(x,t)$.  The wave functions on the left (right) side of the potential well are then given by  $\bm{A}$ ($\bm{B}$). Note that $\bm{A}^{i,o}_l$, and $\bm{B}^{i,o}_l$ where $l=m,n$ represent the probability amplitudes and should not be confused with vector quantities.    
From boundary conditions at the well edges \(x = \pm L/2\) we can then derive the Floquet scattering matrix $\bm {S}$ \cite{blanter2000shot}
\begin{align}\label{eq:S-matrix-def}
    \pmqty{\bm{A}^o_n\\\bm{B}^o_n} =  \sum_{m=-\infty}^{\infty}\bm{S}_{nm}\pmqty{\bm{A}^i_m \\ \bm{B}^i_m},
\end{align}
containing the probability amplitudes of scattering from sideband \(m\) to \(n\) when propagating through the quantum well. 
The  unitary scattering matrix for real currents is given by \cite{blanter2000shot}
\begin{align}\label{eq:S-matrix-comps}
    s(E_n,E_m) &= \pmqty{s_{LL}(E_n,E_m) & s_{LR}(E_n,E_m) \\ s_{RL}(E_n,E_m) & s_{RR}(E_n,E_m)} \\
    &= \sqrt{\frac{\Re(k_{xn})}{\Re(k_{xm})}}\bm{S}_{nm} = \pmqty{r_{nm} & t'_{nm} \\ t_{nm} & r'_{nm}}.\nonumber
\end{align}
Here $k_{xl}$ denotes the Floquet wave vector outside the harmonic well for the $l$-th sideband. We here discard all amplitudes for evanescent non-propagating modes with \(\Re(k_{xl}) = 0\). Generally, we have to consider the incident energy to be above a certain threshold value where negative modes are evanescent, keeping \(m,n \in \mathbb{N} \equiv [0,\infty)\), such that the ${\bm S}$-matrix becomes unitary.
Overall, we read \(s_{\alpha_\beta}(E_n,E_m)\) as the process where a particle enters from side \(\beta\) with energy \(E_m\) and scatters into side \(\alpha\) with energy \(E_n\). The sides can be either \(L\) (left) or \(R\) (right). We can thus identify the elements \(r_{nm},\ t_{nm}\) as the amplitudes for reflecting $(r)$ and transmitting $(t)$ an incoming particle on the left from the $m$-th to the $n$-th sideband. Primed quantities correspond to the backward amplitudes for the scattering. 
In all our calculations we assume a single electron wave incident from left with energy \(E_0 = E_F\) carrying the momentum \(k_{x0}\). This refers to the fact that we only have one incident sideband $m=0$ from the left side.     
The total transmission is then calculated as 
\begin{equation}
        T = \sum_{n=0}^\infty \abs{s_{RL}(E_n,E_0)}^2 = \sum_{n=0}^\infty \abs{t_{n0}}^2.
    \label{eq:transmission-final}
\end{equation}
In order to achieve numerical accuracy with a finite termination of the infinite sum,  the number of  sidebands $n_{\rm min}$ has to be greater than $V_1/\hbar \omega$.  In the next two subsections, we describe the specific wave functions with explicit forms of the Floquet wave vectors for the LBT and QBT dispersions.

%Discussion on shot noise:
The  pumped current becomes zero if there is no electric or temperature bias
between the left and right sides of the well, or electrodes. However, there exist current fluctuations around this zero mean value of current.  
The shot noise measures these current fluctuations and can be considerably large due to virtual transport of electrons and holes during a driving cycle. The pumped shot noise at low temperature is given by \cite{blanter2000shot}
\begin{align}
    \mathcal{N}_{\alpha\beta} =& \frac{e^2}{2h}\int_0^\infty dE \sum_{\gamma\delta}\sum_{mnp\in\mathbb{Z}} M_{\alpha\beta\gamma\delta}(E,E_m,E_n,E_p)\nonumber \\
    &\times[f_0(E_n)-f_0(E_m)]^2
    \label{eq:shotnoise-Nint}
\end{align}
where \(e\) and \(h\) are the fundamental charge and Planck's constant, respectively, and the multi-variate element is
\begin{align}\label{eq:shotnoise-M}
    M_{\alpha\beta\gamma\delta}(E,E_m,E_n,E_p) =& s_{\alpha\gamma}^*(E,E_n)s_{\alpha\delta}(E,E_m)\\
    &\times s_{\beta\delta}^*(E_p,E_m)s_{\beta\gamma}(E_p,E_n).\nonumber
\end{align}
This quantity records the scattering processes contributing to the pumped shot noise, which reads as two particles entering from \(\gamma\) and \(\delta\) electrodes with energies \(E_n\) and \(E_m\), respectively. They are then scattered to the \(\alpha\) and \(\beta\) electrodes with energies \(E\) and \(E_p\), respectively. Here the finite temperature Fermi-Dirac distribution is $f_0(E_n) = (e^{(E_n-E_F)/k_BT}+1)^{-1}$  and we also define \(\mathcal{F}_{nm} = [f_0(E_n)-f_0(E_m)]^2\) for ease of notation. Furthermore, all energies are taken as \(E_l = E+l\hbar\omega\), with the same zeroth mode energy \(E_0 = E\) in all quantities, corresponding to the integration variable. The pumped shot noise, therefore, records the transport of all energy channels below the Fermi level \(E_F\), which we can take as an upper limit of the integral on $E$ instead of $\infty$ in Eq. (\ref{eq:shotnoise-Nint}). 
Moreover, the pumped shot noise is unitary, \(\mathcal{N}_{LL}=-\mathcal{N}_{LR}=-\mathcal{N}_{RL}=\mathcal{N}_{RR}\), meaning we only need to consider one of these quantities and we choose \(\mathcal{N} =\mathcal{N}_{LL} \). Note that \(s(E_n,E_m)\) disregards the evanescent modes as zero, thus we can consider only \(m,n,p\in \mathbb{N}\). We then have
\begin{align*}
    \mathcal{N} =& \frac{e^2}{2h}\int_0^{E_F}  dE \sum_{\gamma\delta}\sum_{mnp\in\mathbb{N}} M_{LL\gamma\delta}(E,E_m,E_n,E_p)  \mathcal{F}_{nm}
\end{align*}

Using the symmetry of \(\mathcal{F}_{nm}\), we take a closer look at the integrand for the pumped shot noise $\mathcal{N}$ to arrive at a more explicit expression. We have 
\begin{align*}
    \sum_{mnp\in\mathbb{N}} &M_{LL\gamma\delta}(E,E_m,E_n,E_p) \mathcal{F}_{nm} = \\
    \sum_{mnp\in\mathbb{N}} & s_{L\gamma}^*(E,E_n)s_{L\delta}(E,E_m)s_{L\delta}^*(E_p,E_m)\\
    &\times s_{L\gamma}(E_p,E_n)\mathcal{F}_{mn}
\end{align*}
By denoting \( (s_{\alpha\beta})_{ij} = s_{\alpha\beta}(E_i, E_j)\) to shift the focus from the \(2\times 2\) matrix \(s(E_i, E_j)\) to the \(n\times m\) matrix \(s_{\alpha\beta}\) the expression becomes
\begin{align*}
    \sum_{mnp\in\mathbb{N}} & (s_{L\gamma}^*)_{0n}(s_{L\delta})_{0m}(s_{L\delta}^*)_{pm} (s_{L\gamma})_{pn}\mathcal{F}_{mn}  \\
    &= \sum_{mn\in\mathbb{N}} (s_{L\gamma}^\dagger)_{n0}(s_{L\delta})_{0m}(s_{L\delta}^\dagger s_{L\gamma})_{mn}\mathcal{F}_{mn}  \\
   &= \sum_{mn\in\mathbb{N}}  (s_{L\delta})_{0m}([s_{L\delta}^\dagger s_{L\gamma}]\odot \mathcal{F})_{mn} (s_{L\gamma}^\dagger)_{n0}  \\
   &=  (s_{L\delta}[[s_{L\delta}^\dagger s_{L\gamma}]\odot \mathcal{F}]s_{L\gamma}^\dagger)_{00}  \\
\end{align*}
where \(\odot\) is the Hadamard product, an element-wise product between two matrices. Including the final summation for the integrand, arrive the final expression shot noise expression
\begin{align}
    \mathcal{N}(E_F) =& \frac{e^2}{2h}\int_0^{E_F}  (r[[r^\dagger r]\odot \mathcal{F}]r^\dagger + r[[r^\dagger t']\odot \mathcal{F}]t'^\dagger \nonumber\\
    &+ t'[[t'^\dagger r]\odot \mathcal{F}]r^\dagger + t'[[t'^\dagger t']\odot \mathcal{F}]t'^\dagger)_{00}dE.
    \label{eq:shotnoise-final}
\end{align}
We can thus analyze the shot noise spectra as a function of the incident energy $E_F$ similar to the transmission spectra obtained in Eq. (\ref{eq:transmission-final}). Already here we can anticipate that the shot noise  is expected to exhibit a distinct behavior when there is a peak or dip in the transmission spectra. In addition, the differential shot noise $\mathcal{N}'=\partial \mathcal{N}(E_F)/ \partial E_F$ can detect also small changes in the shot noise spectra itself. For our analysis, we examine both $\mathcal{N}$ and $\mathcal{N}'$ for the LBT and QBT cases. 

%%%%%%%%%%%%%%%%%%%%%%%%%%%%%%%%%%%%%%%%%%%%%%%%%%%%%%%%%%%%%%%%%%%%
\subsection{LBT}
\label{sec2_1}
%%%%%%%%%%%%%%%%%%%%%%%%%%%%%%%%%%%%%%%%%%%%%%%%%%%%%%%%%%%%%%%%%%%%

\begin{figure}
    \centering
    \includegraphics[width = \linewidth]{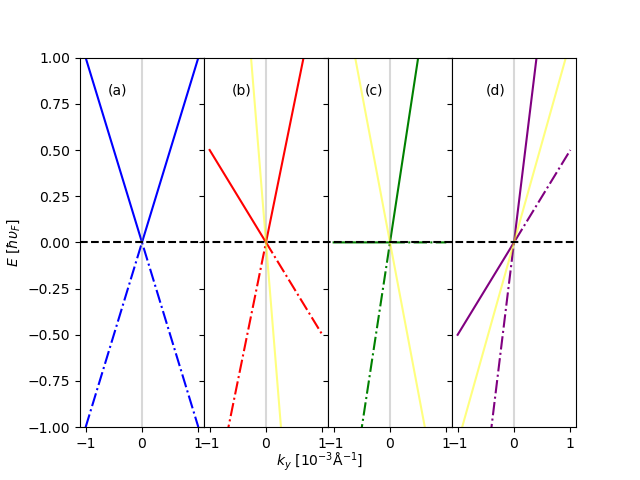}
    \caption{Evolution of LBT dispersion obtained from Eq.(\ref{eq:graphene-energy})  with \(k_x = 0\) for  tilt strengths \(\tau_y/\upsilon_F = 0, 0.5, 1, 1.5\) in (a,b,c,d), respectively. Solid (dashed-dotted) lines correspond to the \(E_+\) (\(E_-\)) bands. Grey line represents the vertical symmetry axis, bifurcating the untiled conical dispersion for  zero tilt \(\tau_y = 0\). Yellow line represents the rotated symmetry axis, bifurcating the tiled conical dispersion for non-zero tilt \(\tau_y \ne  0 \).
}
    \label{fig:graphene-band-tilt}
\end{figure}

To study the LBT case with tilt $n_0(k_y)= \hbar \tau_y k_y$,
we consider $n_x(\bm{k})= \hbar\upsilon_F k_x, n_y(\bm{k})=\hbar \upsilon_F k_y, n_z(\bm{k})=0$. This results in the LBT Hamiltonian
\begin{align} \label{eq:Hamiltonian-graphene}
    H = \hbar\upsilon_F (k_x\sigma_x + k_y\sigma_y) + \hbar \tau_y k_y \sigma_0,
\end{align}
where \(\upsilon_F\) is the Fermi velocity.  Note that $\tau_y=0$, represents the typical graphene dispersion \cite{Neto09} (see Fig. \ref{fig:graphene-band-tilt}(a)), while $\tau_y \ne 0$ corresponds to what would be tilted graphene. The energy spectrum is given by 
\begin{align}\label{eq:graphene-energy}
    E_\pm = \hbar  \tau_y k_y \pm \hbar \upsilon_F \abs{\bm{k}},
\end{align}
with $|{\bm{k}}|=(k_x^2 +k_y^2)^{1/2}$. 
Note that the band dispersion remains gapless at ${\bm{k}}=(k_x,k_y)=(0,0)$ irrespective of the strength of the tilt term.

We systematically demonstrate the evolution of the linear bands
under the application of the tilt in Fig. \ref{fig:graphene-band-tilt}. For $\tau_y =0$ in Fig.~\ref{fig:graphene-band-tilt}(a), the valence (dashed dotted) and conduction (solid) bands are symmetric with respect to the $k_y=0$ axis, depicted by a grey line. 
As $\tau_y$ increases with $\tau_y< \upsilon_F$, the valence and conduction bands become asymmetric with respect to this axis as displayed in Fig.~\ref{fig:graphene-band-tilt}(b). The valence (conduction) band becomes maximally asymmetric for $\tau_y=\upsilon_F$ when the $k_y>0$ ($k_y<0$) part of the valence (conduction) band lying exactly at zero energy as shown in Fig.~\ref{fig:graphene-band-tilt}(c).  For $\tau_y>\upsilon_F$, the $k_y>0$ ($k_y<0$) part of what was priorly the valence (conduction) band now becomes the conduction (valence) band as found in Fig.~\ref{fig:graphene-band-tilt}(d). 
This is due to the fact that $E_{+}$ ($E_{-}$) no longer remains the conduction (valence) band for all values of $k_y$ with $k_x=0$.
One can define an axis, designated by a yellow line in Fig. \ref{fig:graphene-band-tilt}, with respect to which the two branches of the valence ($E<0$) and conduction ($E>0$) bands are symmetric. For the untilted case with $\tau_y=0$, this axis of the conical dispersion is
at right angle $\Theta=\pi/2$ with the $E=0$ axis. As the tilt increases, this axis i.e., the yellow line, moves counter-clockwise. 
Once the tilt passes through the critical tilt  $\tau_y=\upsilon_F$, this axis suddenly makes a clockwise movement due to partial reversal among the valence and conduction bands.  
For the overtilted case with $\tau_y>\upsilon_F$, the 
counter-clockwise motion of the conical axis is again observed. 
For simplicity, we always restrict ourselves to the case with $\tau_y < \upsilon_F$ as the  counter-clockwise movement of the conical axis is always noticed  when the tilt increases. This further corresponds to the fact that $E_{+}>0$ ($E_{-}<0$) always represents the conduction (valence) band for all values of $k_y$ when $k_x=0$. We further consider the physically realistic parameters \cite{zhu2015fano} $\upsilon_F \sim 10^6 {\rm m/s}, \tau_y \sim 10^4 {\rm m/s}$ for our calculations.

%%%%%%%%%%%%%%%%%%%%%%%%%%%%%%%%%%%%%%%%%%%%%%%%%%%%%%%%%%%%%%%

%%%%%%%%%%%%%%%%%%%%%%%%%%%%%%%%%%%%%%%%%%%%%%%%%%%%%%%%%%%%%%%

We can solve the time-independent Schr\"odinger equation to find out the \(x\)-components of the
wave vector $k_x$ and $q$  for the free and potential regions, respectively, resulting in 
\begin{align}\label{eq:tilt-graphene-momentum}
    k_x^2 &= \frac{(E-\tau_y\hbar k_y)^2}{(\hbar\upsilon_F)^2}-k_y^2,~~~~ {\rm for} ~x>|L/2|\nonumber\\
    q^2 &= \frac{(E-V_0-\tau_y\hbar k_y)^2}{(\hbar\upsilon_F)^2}-k_y^2, ~~~~ {\rm for} ~x\le |L/2|.
\end{align}
We can then exploit Floquet theory to also solve the time-dependent Schr\"odinger equation.  
In the Floquet scattering formalism \cite{Wenjun99, zhu2015fano}, we can write the general form of the wave function as
\begin{widetext}
\begin{align}\label{eq:tiltgraphenewf}
    \psi_n(x,t) = e^{-iE_nt/\hbar}\begin{cases}
        \bm{A}_n^i\pmqty{1 \\ s_ne^{i\phi_n}}e^{ik_{xn}x}-\bm{A}_n^o\pmqty{-1 \\ s_ne^{-i\phi_n}}e^{-ik_{xn}x}, &x<-L/2\\
        \displaystyle\sum_{m=-\infty}^{\infty}\bqty{a_m\pmqty{1 \\ s'_me^{i\theta_m}}e^{iq_mx}-b_m\pmqty{-1 \\ s'_me^{-i\theta_m}}e^{-iq_mx}}J_{n-m}\pmqty{\frac{V_1}{\hbar \omega}}, &\abs{x}\leq L/2\\
        -\bm{B}_n^i\pmqty{-1 \\ s_ne^{-i\phi_n}}e^{-ik_{xn}x}+\bm{B}_n^o\pmqty{1 \\ s_ne^{i\phi_n}}e^{ik_{xn}x}, & x>L/2
    \end{cases}
\end{align}
\end{widetext}
where \(\bm{A}_n^{i},\bm{A}_n^{o}\) and \(\bm{B}_n^{i},\bm{B}_n^{o}\) are as before, the probability amplitudes on the left (\(\bm{A}\)) and right (\(\bm{B}\)) side of the potential, respectively, with the superscripts denoting the incoming (i) or outgoing (o) amplitudes, in the \(n\)-th sideband. Inside the potential well, \(a_m\) and \(b_m\) denote the right and left propagating amplitudes in the \(m\)-th sideband, respectively. The other parameters are here \(s_n = \text{sign}(E_n-\tau_y\hbar k_y)\), \(s'_m = \text{sign}(E_m-V_0-\tau_y\hbar k_y)\), \(k_{xn} = \sqrt{[(E_n-\tau_y\hbar k_y)/\hbar\upsilon_F]^2-k_y^2}\), \(q_m = \sqrt{[(E_m-V_0-\tau_y\hbar k_y)/\hbar\upsilon_F]^2-k_y^2}\), \(\phi_n = \arctan(k_y/k_{xn})\), and \(\theta_m = \arctan(k_y/q_m)\), and $J_{n-m}$ is the $n-m$th Bessel function of the first kind. We here consider $n=0$ for a single electron incident from the left as discussed previously. By matching the wave functions in Eq.~\eqref{eq:tiltgraphenewf} at the two well boundaries, we can derive the explicit form of $\bm{S}$-matrix in Eqs.~(\ref{eq:S-matrix-def}-\ref{eq:S-matrix-comps}), see Appendix \ref{appA} for details. 

Having discussed the propagating modes in Eq.~(\ref{eq:tiltgraphenewf}), we now turn our attention to the quasi-bound states inside the static potential well \cite{zhu2015fano}. Their energies result in imaginary (real) wave-vectors outside (inside) the potential well such that these quasi-bound states  decay exponentially (propagate without decay)  outside (inside) the potential well along the $\pm x$-direction.   Note that the electrons still propagate freely along the $y$-direction. 
To investigate the quasi-bound states, we have to consider the wave function as given in  Eq.~(\ref{eq:tiltgraphenewf})  with \(n = m = V_1 = 0\). From there  we can find energy roots of the secular equation arising from the boundary conditions at $x=\pm L/2$ associated with the appropriate wave functions \cite{zhu2015fano}. 
This amounts to energy solutions of the quasi-bound states as follows
\begin{align}\label{eq:LBT-bound}
     e^{-2iq_0L} = \frac{2-s_0's_0\cos(\phi_0-\theta_0)}{2+s_0's_0\cos(\phi_0+\theta_0)}
\end{align}
Note that this is a transcendental equation and an analytical closed form is not possible to obtain. Still, we can obtain numerical solutions, corresponding to multiple energies $E_b$ of the  quasi-bound states, through the graphical method.  
Moreover, given the real-valued nature of  $\theta_0$ and $\phi_0$, we can actually obtain analytical bounds on the quasi-bound state energies $E_b$ as follows: $ \hbar k_y( \upsilon_F+\tau_y) - |V_0| \le E_b  \le \hbar k_y( \upsilon_F+\tau_y)$. We can also define 
\(E_b = \mathcal{E}_b+\tau_y\hbar k_y\) 
such that $ \hbar k_y\upsilon_F - |V_0| \le \mathcal{E}_b  \le \hbar k_y\upsilon_F $,
where $\mathcal{E}_b$ designates the energy of the quasi-bound state in the absence of tilt.
This suggests that the tilt simply shifts the energy of the quasi-bound state depending on the sign of 
$\tau_y k_y$ term.  Importantly, for the driven case when the wave-vector $q_{l}$, associated with the Floquet sideband, matches with the wave-vector corresponding to the quasi-bound state within the potential well, we expect a resonance in the transmission spectra. In other words, when the Floquet sideband energy $E_l$ is identical with the  quasi-bound state energy $E_b$ i.e., $E_F + l \hbar \omega = E_b$, we expect to   observe a Fano resonance.  As a consequence, we expect this same linear shift given by $\tau_y\hbar k_y$ in the quasi-bond energy to also appear in the transmission spectra with the Fano resonant energies separated by \(\tau_y\hbar k_y\) for different values of $\tau_y$ or $k_y$.

%%%%%%%%%%%%%%%%%%%%%%%%%%%%%%%%%%%%%%%%%%%%%%%%%%%%%%%%%%%%%%%%%%%%
\subsection{QBT}
\label{sec2_2}
%%%%%%%%%%%%%%%%%%%%%%%%%%%%%%%%%%%%%%%%%%%%%%%%%%%%%%%%%%%%%%%%%%%%

To study the QBT case with tilt $n_0(k_y)=\hbar  \tau_y k_y$, 
we consider $n_x(\bm{k})=\hbar^2 k_x k_y/\mu, n_y(\bm{k})=0, n_z(\bm{k})= \hbar^2 (k_y^2 -k_x^2)/(2 \mu)$. This results in the QBT Hamiltonian \cite{bera2021floquet}
\begin{align}\label{eq:Hamiltonian-QBT}
    H =& \frac{\hbar^2}{2\mu}(2k_xk_y\sigma_x+(k_y^2-k_x^2)\sigma_z)+\hbar  \tau_y k_y \sigma_0
\end{align}
where  $\mu$ denotes the effective mass. 
Here the energies take the form
\begin{equation}
E_\pm = \hbar  \tau_y k_y \pm \frac{\hbar^2}{2\mu}|{\bm{k}}|^2,
\label{eq:qbt-energy}
\end{equation}
where we note that the parabolic band dispersion remains gapless at ${\bm{k}}=(k_x,k_y)=(0,0)$ even after the introduction of the tilt term. 

%%%%%%%%%%%%%%%%%%%%%%%%%%%%%%%%%%%%%%%%%%%%%%%%%%%%%%%%%%%%%%%
\begin{figure}
    \centering
    \includegraphics[width=\linewidth]{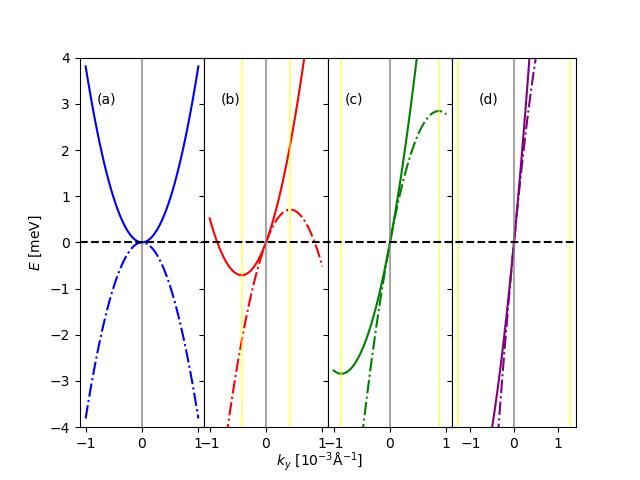}
    \caption{Evolution of QBT dispersion obtained from Eq.~(\ref{eq:qbt-energy}) with \(k_x = 0\) for  tilt strengths  \({\tau_y} = 0, 0.5, 1, 1.5\) in units of \(10^6\)m/s in (a,b,c,d), respectively. Solid (dashed-dotted) lines correspond to the \(E_+\) (\(E_-\)) bands. Grey line represents the vertical symmetry axis, bifurcating the undisplaced parabolic energy dispersion for zero tilt \(\tau_y = 0\). Yellow lines represent the vertical symmetry axes bifurcating the displaced parabolic energy dispersion for non-zero tilt \(\tau_y \ne 0 \).
}
    \label{fig:qbt-band-tilt}
\end{figure}

%%%%%%%%%%%%%%%%%%%%%%%%%%%%%%%%%%%%%%%%%%%%%%%%%%%%%%%%%%%%%%%

We demonstrate the evolution of the QBT dispersion with increasing tilt strength in Fig.~\ref{fig:qbt-band-tilt}. In the untilted case $\tau_y=0$ in Fig. \ref{fig:qbt-band-tilt}(a) the parabolic valence and \textcolor{black}{conduction} bands are symmetric around the vertical axis at $k_y=0$, represented by a grey line. The bottom (top) of the conduction (valence) band is at $E=0$, and $(k_x,k_y)=(0,0)$.  The part of the prior conduction (valence) band around the band bottom (top) becomes the new valence (conduction) band immediately after the application of the tilt $\tau_y\ne 0$.
This is due to the fact that $E_{+}$ ($E_{-}$) no longer remains conduction (valence) band for all values of $k_y$ with $k_x=0$ and $\tau_y \ne 0$.
The parabolic conduction (valence) bands instead become symmetric around the vertical axis $k_y=-\mu \tau_y /\hbar$ ($k_y=\mu \tau_y /\hbar$), while the energy at the bottom (top) of the parabolic conduction (valence) band is found to be $-\mu \tau_y^2/2$ ($+\mu \tau_y^2/2$), both represented by yellow lines. This behavior is consistently observed irrespective of the values of $\tau_y$ as shown in Figs. \ref{fig:qbt-band-tilt}(b,c,d). Unlike the tilted LBT shown in Fig. \ref{fig:graphene-band-tilt}, we cannot define a single parabolic axis through $E=0$ and $(k_x,k_y)=(0,0)$ as the 
tilt is linear and dispersion is quadratic with respect to $k_y$. We can think of the two parabolic bands apparently sliding through each other, keeping the meeting point fixed at $E=0$ and $(k_x,k_y)=(0,0)$. 
We obtain an expression for the circle of paraboloid intersection with the zero energy plane as
\begin{align*}
    k_x^2 +\pmqty{k_y\pm \frac{\mu\tau_y}{\hbar}}^2 = \pmqty{\frac{\mu\tau_y}{\hbar}}^2,
\end{align*}
where \(\pm\) is originated from \(E_\mp\).
We here consider the physically realistic parameters \cite{bera2021floquet}$\tau_y \sim 10^4-10^6 {\rm m/s}$, and \(\mu = 0.001 m_e\) with \(m_e\) the free electron mass, for our calculations. 

Just as in the LBT case, we obtain the wave-vectors for the time-independent problem as 
\begin{align} \label{eq:tilt-qbt-momentum}
    k_x^2 &= \frac{2\mu}{\hbar^2}(E-\hbar \tau_y k_y)-k_y^2,~~~~~{\rm for}~x>|L/2|\nonumber \\
    q^2 &= \frac{2\mu}{\hbar^2}\abs{E-V_0-\hbar \tau_y k_y}-k_y^2,~~~~~{\rm for}~x\le |L/2|
\end{align}
for free and potential regions, respectively.  Using a similar Floquet treatment as for the LBT case, we write the wave function as  \cite{bera2021floquet}
\begin{widetext}
\begin{align}\label{eq:tiltwf}
\psi_n(x,y,t) = e^{ik_yy}\begin{cases}
        \eta_{1,n}\pqty{\bm{A}^i_n\pmqty{1 \\ \displaystyle\frac{k_{xn}}{k_y}}e^{ik_{xn}x}+\bm{A}^o_n\pmqty{1 \\ -\displaystyle\frac{k_{xn}}{k_y}}e^{-ik_{xn}x}}\Theta_n^+ & x < -L/2 \\
        \quad \quad +\eta_{4,n}\left(\bm{A}^i_n\pmqty{1 \\ -\displaystyle\frac{k_y}{k_{xn}}}e^{ik_{xn}x}+\bm{A}^o_n\pmqty{1 \\ \displaystyle\frac{k_y}{k_{xn}}}e^{-ik_{xn}x}\right) \Theta_n^-
        \\
        \displaystyle\sum_{m=-\infty}^\infty\left[\eta_{2,m}\left(a_m\pmqty{1 \\ \displaystyle\frac{q_m}{k_y}}e^{iq_mx}+b_m\pmqty{1 \\ -\displaystyle\frac{q_m}{k_y}}e^{-iq_mx}\right)\Theta_m^+\right. &\abs{x} \leq L/2\\
        \left.\quad \quad +\eta_{3,m}\left(a_m\pmqty{1 \\ -\displaystyle\frac{k_y}{q_m}}e^{iq_mx}+b_m\pmqty{1 \\ \displaystyle\frac{k_y}{q_m}}e^{-iq_mx}\right)\Theta_m^-\right]J_{n-m}\pmqty{\frac{V_1}{\hbar \omega}}\\
        \eta_{1,n}\pqty{\bm{B}^o_n\pmqty{1 \\ \displaystyle\frac{k_{xn}}{k_y}}e^{ik_{xn}x}+\bm{B}^i_n\pmqty{1 \\ -\displaystyle\frac{k_{xn}}{k_y}}e^{-ik_{xn}x}} \Theta_n^+ &x>L/2 \\
        \quad \quad +\eta_{4,n}\left(\bm{B}^o_n\pmqty{1 \\ -\displaystyle\frac{k_y}{k_{xn}}}e^{ik_{xn}x}+\bm{B}^i_n\pmqty{1 \\ \displaystyle\frac{k_y}{k_{xn}}}e^{-ik_{xn}x}\right) \Theta_n^-
    \end{cases}
\end{align}
where we use
\begin{align*}
    \eta_{1,n} &= \frac{\abs{k_y}}{\sqrt{k_{xn}^2+k_y^2}},\quad \eta_{2,m} = \frac{\abs{k_y}}{\sqrt{q_m^2+k_y^2}},\quad \eta_{3,m} = \frac{\abs{q_m}}{\sqrt{q_m^2+k_y^2}},\quad
     \eta_{4,n} = \frac{\abs{k_x}}{\sqrt{k_x^2+k_y^2}},\quad k_{xn}^2 = \frac{2\mu}{\hbar^2}(E_n-\hbar \tau_y k_y)-k_y^2\\
    q_m^2 &= \frac{2\mu}{\hbar^2}|E_m-V_0-\hbar \tau_y k_y|-k_y^2,\quad
    \Theta_{m}^\pm = \Theta(\pm[E_{m}-V_0-\tau_y\hbar k_y]), \quad \Theta_{n}^\pm = \Theta(\pm[E_{n}-\tau_y\hbar k_y]).
\end{align*}
\end{widetext}
Here the Heaviside function \(\Theta(z)\) takes care of both the conduction and valence bands  for free and potential regions.
Just as in the LBT case, we derive the explicit form of the ${\bm S}$-matrix as represented in Eqs.~(\ref{eq:S-matrix-def}-\ref{eq:S-matrix-comps}) in Appendix \ref{appB} by exploiting boundary conditions appropriately. 

In order to obtain the energies of the quasi-bound states, we consider \(n=m=V_1 = 0\) as already discussed for the LBT case. By employing boundary conditions at the potential well edges, we can derive the following  transcendental equation from which energy roots $E_b$ can be obtained \cite{bera2021floquet}
\begin{align}\label{eq:QBT-bound}
   \tan(L\sqrt{\frac{2\mu(E_b-\tau_y\hbar k_y -V_0)}{\hbar^2}-k_y^2}) =\\
   \frac{\hbar^2\sqrt{k_y^2-\frac{2\mu (E_b-\tau_y\hbar k_y)}{\hbar^2}}\sqrt{\frac{2\mu (E_b-\tau_y\hbar k_y-V_0)-k_y^2}{\hbar^2}}}{2\mu(E_b-\tau_y\hbar k_y)-\mu V_0 -\hbar^2k_y^2}.\nonumber 
\end{align}
The closed analytical  form of $E_b$ is hard to obtain, so we again resort to multiple graphical solutions. 
Considering the quantities to be real inside the square root, we find  that quasi-bound
states are energetically bounded between $ \hbar^2 k_y^2/(2 \mu) - |V_0| \le \mathcal{E}_b  \le \hbar^2 k_y^2/(2 \mu)$, 
where \( \mathcal{E}_b = E_b- \tau_y\hbar k_y\) denotes the 
quasi-bound energy in the absence of tilt $\tau_y=0$. In particular, we note that the energies of quasi-bound states are modified due to a tilt in an identical manner for the LBT and QBT cases. We hence expect that the Fano resonances become shifted by \(\tau_y\hbar k_y\) regardless of dispersion type.

%%%%%%%%%%%%%%%%%%%%%%%%%%%%%%%%%%%%%%%%%%%%%%%%%%%%%%%%%%%%%%%%%%%%
\section{Results}
\label{sec3}
%%%%%%%%%%%%%%%%%%%%%%%%%%%%%%%%%%%%%%%%%%%%%%%%%%%%%%%%%%%%%%%%%%%%

\begin{figure}
    \centering
    \includegraphics[width=\linewidth]{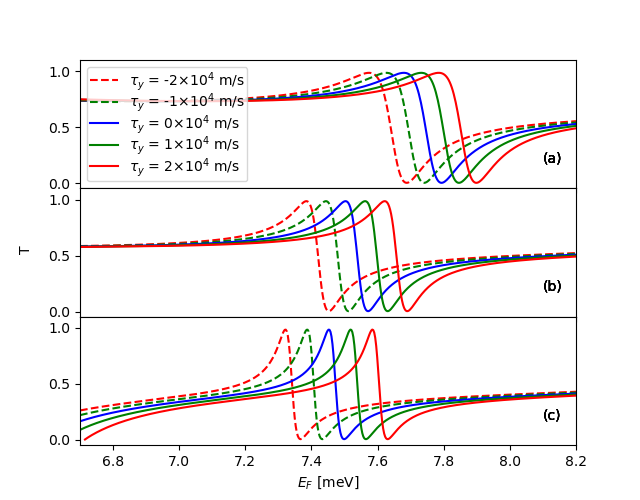}
    \caption{Total transmission spectra obtained from Eq.~(\ref{eq:transmission-final}) for LBT case given by Eq. (\ref{eq:Hamiltonian-graphene})  with 
    transverse momentum  \(k_y = 0.0008\), \(0.0009\)  and \(0.001\) Å\(^{-1}\) in  (a,b,c), respectively for different tilt strengths $\tau_y$. The Fano resonance energy, associated with the peak-dip structure, shifts with changing the tilt strength.   Other parameters: \(L=3000\) Å, \(V_0 = -50\) meV, \(V_1 = 1\) meV, \(\hbar\omega = 4\) meV, \(N = 2\), \(\upsilon_F = 10^6\) m/s.
}
    \label{fig:graphene-transmission-posky}
\end{figure}

\begin{figure}
    \centering
    \includegraphics[width=\linewidth]{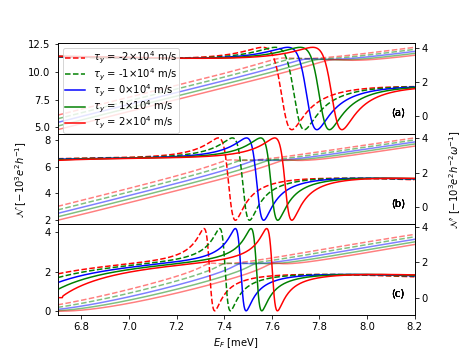}
    \caption{Shot noise spectra ${\mathcal N}$ (\textcolor{black}{light-colored lines}, left axis) and differential shot noise spectra ${\mathcal N}'$ (\textcolor{black}{dark-colored} lines, right axis) obtained from Eq.~(\ref{eq:shotnoise-final}) for LBT case  with  transverse momentum  \(k_y = 0.0008\), \(0.0009\) and \(0.001\) Å\(^{-1}\) in  (a,b,c), respectively for different tilt strengths $\tau_y$. 
 The slope in ${\mathcal N}$ changes, while undulations, namely a peak-dip profile, are observed in ${\mathcal N}'$ around the Fano resonance energy, which shifts with changing tilt strength. Other parameters same as Fig.~\ref{fig:graphene-transmission-posky}.
}
    \label{fig:graphene-shotnoise-posky}
\end{figure}

\begin{figure}
    \centering
    \includegraphics[width=\linewidth]{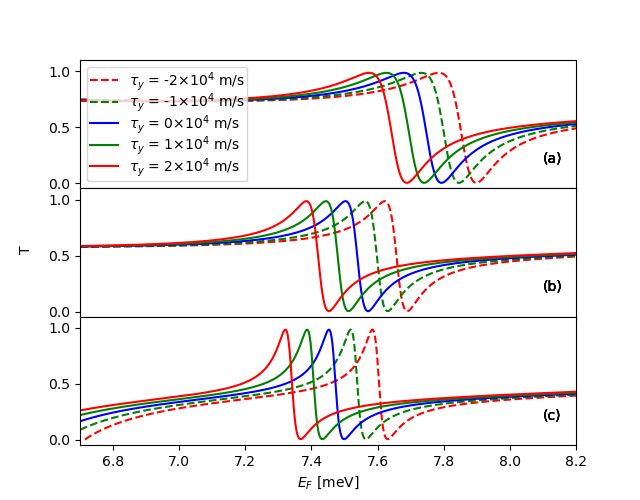}
    \caption{Same as Fig.~\ref{fig:graphene-transmission-posky} but for negative transverse momentum \(k_y = -0.0008\), \(-0.0009\) and \(-0.001\) Å\(^{-1}\) in plots (a,b,c), respectively.
}
    \label{fig:graphene-transmission-negky}
\end{figure}

\begin{figure}
    \centering
    \includegraphics[width=\linewidth]{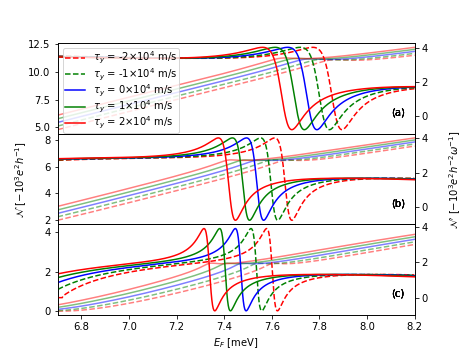}
    \caption{Same as Fig.~\ref{fig:graphene-shotnoise-posky} but for negative transverse momentum \(k_y = -0.0008\), \(-0.0009\) and \(-0.001\) Å\(^{-1}\) in plots (a,b,c), respectively.    
%    Pumped shot noise and its energy derivative of Fig.[\ref{fig:graphene-transmission-negky}]. Superposing like Fig.[\ref{fig:graphene-shotnoise-posky}]
}
    \label{fig:graphene-shotnoise-negky}
\end{figure}

\begin{figure}
    \centering
    \includegraphics[width=\linewidth]{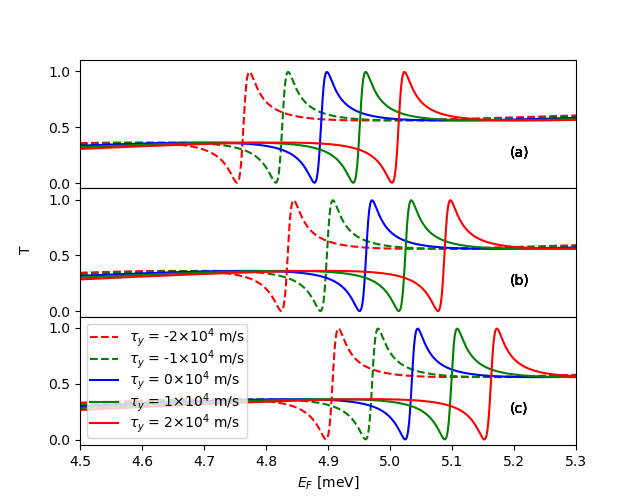}
    \caption{Total transmission spectra obtained from Eq.~(\ref{eq:transmission-final}) for QBT case given by Eq.~(\ref{eq:Hamiltonian-QBT})  with 
    transverse momentum  \(k_y = 0.00095\), \(0.00096\) and \(0.00097\) Å\(^{-1}\) in  (a,b,c), respectively, for different tilt strengths $\tau_y$. The Fano resonance energy, associated with the peak-dip structure, shifts with changing tilt strength. Other parameters: \(L=3000\) Å, \(V_0 = -10\) meV, \(V_1 = 1\) meV, \(\hbar\omega = 4\) meV, \(N = 2\), \(\mu = 0.001m_e\). 
}
    \label{fig:qbt-transmission-posky}
\end{figure}

\begin{figure}
    \centering
    \includegraphics[width=\linewidth]{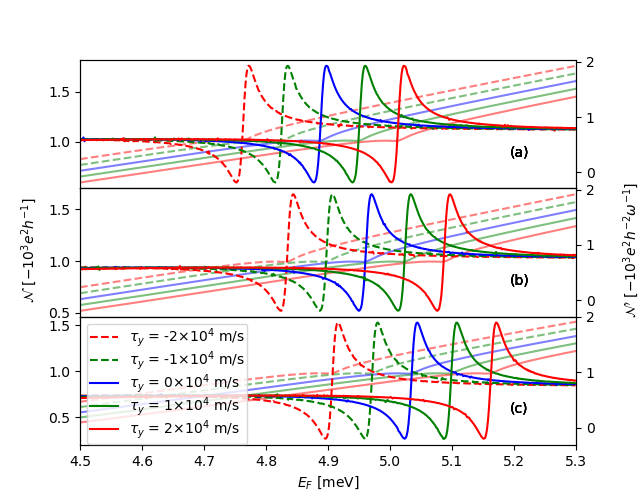}
    \caption{Shot noise spectra ${\mathcal N}$ (\textcolor{black}{light-colored lines}, left axis) and differential shot noise spectra ${\mathcal N}'$ (\textcolor{black}{dark-colored lines}, right axis) obtained from Eq.~(\ref{eq:shotnoise-final}) for QBT case  with  transverse momentum  \(k_y = 0.00095\), \(0.00096\) and \(0.00097\) Å\(^{-1}\) in  (a,b,c), respectively. 
    The slope in ${\mathcal N}$ changes, while undulations are observed in ${\mathcal N}'$ around the Fano resonance energy, which shifts with changing tilt strength. Other parameters same as Fig.~\ref{fig:qbt-transmission-posky}.
}
    \label{fig:qbt-shotnoise-posky}
\end{figure}

\begin{figure}
    \centering
    \includegraphics[width=\linewidth]{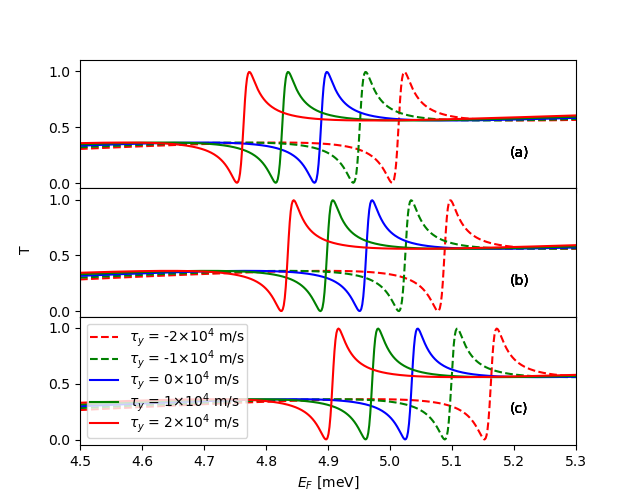}
    \caption{Same as Fig.~\ref{fig:qbt-transmission-posky} but
    for negative transverse momentum \(k_y = -0.00095\), \(-0.00096\) and \(-0.00097\) Å\(^{-1}\) in (a,b,c), respectively.
}
    \label{fig:qbt-transmission-negky}
\end{figure}

\begin{figure}
    \centering
    \includegraphics[width=\linewidth]{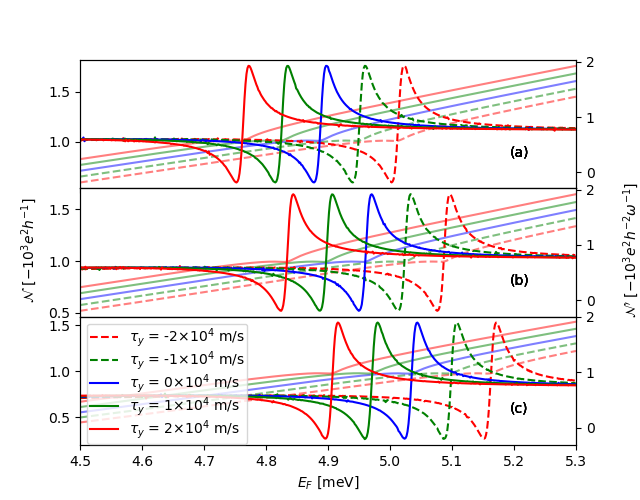}
    \caption{Same as Fig.~\ref{fig:qbt-shotnoise-posky} but
    for negative transverse momentum \(k_y = -0.00095\), \(-0.00096\) and \(-0.00097\) Å\(^{-1}\) in (a,b,c), respectively.
}
    \label{fig:qbt-shotnoise-negky}
\end{figure}

Before  reporting our results explicitly, we first briefly describe the pictorial representation of the findings. 
We show the transmission spectra $T$, obtained from Eq.~(\ref{eq:transmission-final}), with the incident energy $E_F$ for LBT in Figs.~\ref{fig:graphene-transmission-posky} and \ref{fig:graphene-transmission-negky} for positive $k_y>0$ and negative $k_y<0$ transverse momenta, respectively. We repeat the above analysis for QBT in Figs.~\ref{fig:qbt-transmission-posky} and \ref{fig:qbt-transmission-negky}. We scrutinize the shot noise spectra $\mathcal{N}$, computed from Eq. (\ref{eq:shotnoise-final}),
with the incident energy $E_F$ for LBT (QBT) case with positive $k_y>0$ and negative $k_y<0$ transverse motions in the Figs. \ref{fig:graphene-shotnoise-posky} and \ref{fig:graphene-shotnoise-negky} (Figs. \ref{fig:qbt-shotnoise-posky} and  \ref{fig:qbt-shotnoise-negky}), respectively.  

%%%%%%%%%%%%%%%%%%%%%%%%%%%%%%%%%%%%%%%%%%%%%%%%%%%%%%%%%%%%%%%
%%%%%%%%%%%%%%%%%%%%%%%%%%%%%%%%%%%%%%%%%%%%%%%%%%%%%%%%%%%%%%%

%%%%%%%%%%%%%%%%%%%%%%%%%%%%%%%%%%%%%%%%%%%%%%%%%%%%%%%%%%%%%%%
%%%%%%%%%%%%%%%%%%%%%%%%%%%%%%%%%%%%%%%%%%%%%%%%%%%%%%%%%%%%%%%

We start with systematically investigating the transmission spectra for the LBT and QBT cases in Fig.~\ref{fig:graphene-transmission-posky} and Fig.~\ref{fig:qbt-transmission-posky}, respectively. For the LBT (QBT) case we observe a peak (dip) followed by a dip (peak) as a function of $E_F$,  which we attribute to a Fano resonance at $E_F=E^{\rm Fano}_F$.  Generally, a Fano resonance phenomenon is comprised of a combined peak-dip profile,  such that $E^{\rm Fano}_F= E_b \pm n \hbar \omega$ \cite{Wenjun99}.
For a harmonically driven potential well, we can think of this peak-dip (dip-peak) profile in the following qualitative way: The incident electron emits (absorbs) an energy quantum of $\hbar \omega$ to meet with the bound state, while the electron absorbs (emits) an energy quantum of $\hbar \omega$ when leaving the potential well. The above resonance condition, yielding the combined peak-dip profile, is thus satisfied by both $E^{\rm Fano}_F= E_b +n \hbar \omega >0$ and $E^{\rm Fano}_F= E_b - n \hbar \omega >0$. There can be multiple values of $E_b$ for the several quasi-bound states, as well as $n\ge 1$. We can hence often find repeated Fano resonances  in the transmission spectra if we were to examine an extensive range of incident energies. 
In the present case, we find  a single $E_b<0$, computed numerically from Eqs.~(\ref{eq:LBT-bound}) and (\ref{eq:QBT-bound}) with $V_0<0$, resulting in the Fano resonance condition $E^{\rm Fano}_F= E_b +n \hbar \omega$ to be satisfied for certain values of $n$. Overall we find numerical agreement between the position of the Fano resonance and the numerically obtained values for $E_b$.

In addition, we notice that for positive transverse motion in the LBT case, the Fano resonances appear at lower values of $E_F$ with increasing $k_y$ as depicted in Fig. \ref{fig:graphene-transmission-posky}. The same trend is observed for negative transverse motion in Fig. \ref{fig:graphene-transmission-negky} with increasing $|k_y|$. On the other hand, Fano resonances appear at higher values of $E_F$ with increasing $k_y$ ($|k_y|$) for QBT case as observed in Fig. \ref{fig:qbt-transmission-posky}  (Fig. \ref{fig:qbt-transmission-negky}). Therefore, we conclude that the transverse motion $k_y$
is connected with the Fano resonance in terms of the energies of the quasi-bound states. The incident energy decreases (increases) with increasing the magnitude of the  transverse
motion for the LBT (QBT) case, resulting in a red (blue) shift in the $E^{\rm Fano}_F$. 
Moreover, we also find that the peak and dip structure becomes more prominent with increasing $|k_y|$ for LBT case. By contrast, the peak and dip structures are almost unaltered with increasing  $|k_y|$ in the QBT case.  We have confirmed that our results for the untilted cases agree with previous findings \cite{zhu2015fano, bera2021floquet}.

%%%%%%%%%%%%%%%%%%%%%%%%%%%%%%%%%%%%%%%%%%%%%%%%%%%%%%%%%%%%%%%
%%%%%%%%%%%%%%%%%%%%%%%%%%%%%%%%%%%%%%%%%%%%%%%%%%%%%%%%%%%%%%%

%%%%%%%%%%%%%%%%%%%%%%%%%%%%%%%%%%%%%%%%%%%%%%%%%%%%%%%%%%%%%%%
%%%%%%%%%%%%%%%%%%%%%%%%%%%%%%%%%%%%%%%%%%%%%%%%%%%%%%%%%%%%%%%

We also investigate the evolution of the $E^{\rm Fano}_F$ by changing $\tau_y$. For positive transverse motion $k_y>0$, the Fano resonance occurs at higher (lower) energies with increasing positive (negative) tilt i.e., $\tau_y >0$ ($\tau_y <0$) in both the cases of LBT and QBT as shown in Figs.~\ref{fig:graphene-transmission-posky} and  \ref{fig:qbt-transmission-posky}. On the other hand, for negative transverse motion $k_y<0$, an exactly opposite trend is observed for both LBT and QBT cases, where the Fano resonance occurs at higher (lower) energies with increasing negative (positive) tilt, see Figs. \ref{fig:graphene-transmission-negky} and  \ref{fig:qbt-transmission-negky}. Importantly, $E^{\rm Fano}_F$ for the untilted case $\tau_y=0$ always appear in the middle between the other $E^{\rm Fano}_F$'s for positive and negative tilt as $\tau_y$ values chosen are symmetric around $\tau_y=0$. As a consequence, the sign of the product $\tau_y k_y$ is determine the direction of the energy shift in the  $E^{\rm Fano}_F$. To be precise, Fano resonances shift to higher (lower) energy values when $\tau_y k_y$ is positive (negative).  This is in complete agreement with our analytical analysis of the energy of the quasi-bound state \(  E_b= \mathcal{E}_b + \tau_y\hbar k_y\) in Sec.~\ref{sec2}.

%%%%%%%%%%%%%%%%%%%%%%%%%%%%%%%%%%%%%%%%%%%%%%%%%%%%%%%%%%%%%%%
%%%%%%%%%%%%%%%%%%%%%%%%%%%%%%%%%%%%%%%%%%%%%%%%%%%%%%%%%%%%%%%

We next examine the pumped shot noise spectra $\mathcal{N}$ with the incident energy $E_F$. We find that the shot noise profile changes its slope around the Fano resonance energy $E_F^{\rm Fano}$, see \textcolor{black}{light-colored} lines and left axis in Figs.~\ref{fig:graphene-shotnoise-posky},  \ref{fig:graphene-shotnoise-negky},  \ref{fig:qbt-shotnoise-posky}, and  \ref{fig:qbt-shotnoise-negky}). This  change in profile can be regarded as an inflection region. Notably then, the differential shot noise 
spectra $\mathcal{N}'$ at incident energy  $E_F$ is able to identify the Fano resonance energy $E_F^{\rm Fano}$ more clearly, see \textcolor{black}{dark-colored} lines and right axis in Figs.~\ref{fig:graphene-shotnoise-posky},  \ref{fig:graphene-shotnoise-negky},  \ref{fig:qbt-shotnoise-posky}, and  \ref{fig:qbt-shotnoise-negky}). More precisely, the differential shot noise carries an undulation, comprised of dip and peak structures, around $E_F=E_F^{\rm Fano}$ which indicates the occurrence of the Fano resonance.  
Upon a detailed inspection, we find that all the relevant Floquet sidebands contribute to the 
shot noise,  while the contribution of a single dominant Floquet channel is more apparent in the differential shot noise. 
We observe the same behavior of the shot noise and its derivative for both the  positive and negative transverse motions, as well as positive and negative tilt. In other words, the differential shot noise 
spectra qualitatively follows the transmission spectra, such that the effects of the tilts are equivalently captured in LBT and QBT cases.
Interestingly, the inflection region in the shot noise spectra reduces for LBT case when the transverse motion $|k_y|$ increases, resulting in sharp features in the differential shot noise. By contrast, the inflection region does not change when $|k_y|$ increases for the QBT case, leading to the more unaltered features in the  differential shot noise. This is also in agreement with the changing transmission peak-dip features between LBT and QBT cases.

%%%%%%%%%%%%%%%%%%%%%%%%%%%%%%%%%%%%%%%%%%%%%%%%%%%%%%%%%%%%%%%
%%%%%%%%%%%%%%%%%%%%%%%%%%%%%%%%%%%%%%%%%%%%%%%%%%%%%%%%%%%%%%%

%%%%%%%%%%%%%%%%%%%%%%%%%%%%%%%%%%%%%%%%%%%%%%%%%%%%%%%%%%%%%%%
%%%%%%%%%%%%%%%%%%%%%%%%%%%%%%%%%%%%%%%%%%%%%%%%%%%%%%%%%%%%%%%

%%%%%%%%%%%%%%%%%%%%%%%%%%%%%%%%%%%%%%%%%%%%%%%%%%%%%%%%%%%%%%%

%%%%%%%%%%%%%%%%%%%%%%%%%%%%%%%%%%%%%%%%%%%%%%%%%%%%%%%%%%%%%%%

%%%%%%%%%%%%%%%%%%%%%%%%%%%%%%%%%%%%%%%%%%%%%%%%%%%%%%%%%%%%%%%%%%%%
\section{Discussion}
\label{sec4}
%%%%%%%%%%%%%%%%%%%%%%%%%%%%%%%%%%%%%%%%%%%%%%%%%%%%%%%%%%%%%%%%%%%%
Having reported our results in the previous section, already pointing out several features, we here further discuss the results.
Importantly, given the fact that we find that the Fano resonance energies  $E_F^{\rm Fano}$ are captured by the peak-dip structures in the transmission as well as differential shot noise spectra, we can extract information about the tilt strength \(\tau_y\) by investigating the positive $k_y>0$ and negative $k_y<0$ transverse motion. Doing so, we define a tilt function  
\[
\mathcal{C}(\tau_y, k_y) = \frac{E^{\rm Fano}_F(k_y,\tau_y)-E^{\rm Fano}_F(-k_y,\tau_y)}{2\hbar k_y}
\]
that can accurately capture the tilt strength. 
Combining this with the general condition for the Fano resonance, \(E^{\rm Fano}_F(k_y,\tau_y) = E_b(k_y,\tau_y) + n\hbar \omega\) and \(  E_b(k_y,\tau_y)= \mathcal{E}_b(k_y,0) + \tau_y\hbar k_y\), we obtain 
$\mathcal{C}(\tau_y,k_y)=\tau_y$  for $\tau_y \ne 0$ and simply $\mathcal{C}(\tau_y,k_y)=0$ for untilted dispersions with $\tau_y=0$.
The overall reason for this behavior is that the Fano resonances, obtained in the tilted system, appear symmetrically with respect to that for the untilted system as far as the sign of $k_y$ is concerned. Thus, defining $\Delta(k_y,\tau_y)=E_b(k_y,\tau_y)-\mathcal{E}_b(k_y,0) $, we find that 
$\Delta(k_y,\tau_y) \to -\Delta(-k_y,\tau_y) $, irrespective of the sign of $\tau_y$. 
Therefore, without any prior knowledge of the tilt strength, just identifying the Fano resonance peaks for positive and negative transverse motion we can estimate the strength of the tilt from the tilt function $\mathcal{C}(\tau_y, k_y)$. For both the LBT and QBT cases, the untilted band dispersions remain symmetric under the sign reversal of $k_y$ i.e., $E_{\pm}(k_y) \to E_{\pm}(-k_y)$. On the other hand, for the tilted case, the energy of the bands change its sign under $k_y \to -k_y$ i.e., $E_{\pm}(k_y) \to -E_{\mp}(-k_y)$. The above property in the energy dispersion acts as a backbone behind the observation of Fano resonance under the sign change of the product $\tau_y k_y$.

In addition, we find an increase and decrease in the $E_F^{\rm Fano}$ with increasing the magnitude of the transverse motion for QBT and LBT case, respectively. Moreover, the tilt
modifies the  $E_F^{\rm Fano}$ in a different manner for QBT case as compared to LBT case. Therefore, the intrinsic nature of the band dispersion plays important role in the occurrence of Fano resonance  for different values of transverse momentum for a given tilt strength. Thus the properties of the Fano resonance can be used to distinguish between LBT and QBT dispersion.
Finally we note that, for a given value of $k_y$, the change in the  $E_F^{\rm Fano}$ with positive and negative tilt is entirely dependent on the specific form of the tilt term. In  both the cases of LBT and QBT, we consider the same linear tilt terms that result in  the shift in the $E_F^{\rm Fano}$ depending on the sign of  $\tau_y k_y$. However, a non-linear tilt proportional to $\tau_y k_y^n$ with $n>1$ could change the above correspondence.

\textcolor{black}{Having discussed transverse tilt along the $y$-direction and with transport along the $x$-direction, we also briefly discuss the validity of our results in presence of a more generic tilt term with a component also along the transport direction.
Once we allow for a  finite  tilt amplitude along the potential direction i.e. the $x$-direction, the situation changes non-trivially. The wave-vector along the propagation direction now depends on the above tilt strength. Given all other parameters fixed, 
we find multiple possible solutions for the wave-vector with a definite sign which is in contrast to the $y$-tilted case, where only one such possibility arises. To be more precise, in the case of a longitudinal tilt component, the conditions for obtaining the positive and negative  wave-vectors, equivalent to the expressions derived in Eqs.~(\ref{eq:tilt-graphene-momentum}) and (\ref{eq:tilt-qbt-momentum}) for the transverse tilt, are very complicated as they are intertwined with the energy conditions.
We expect that the Fano resonances will be taking place as usual, while the bound state energy is modified by the longitudinal tilt in addition to the transverse tilt. Overall this will cause additional Fano resonances to appear.
For a given strength of the longitudinal tilt, there will be satellite Fano resonances, occurring around the already existed transverse tilt-mediated Fano resonances that we have thoroughly already report on in the previous sections. These new Fano resonances, occurring for positive and negative values of $k_y$,  will be related to each other by a far more complicated relationship as compared to that for the  transverse tilt-mediated Fano resonances given by $\Delta(k_y,\tau_y) \to -\Delta(-k_y,\tau_y) $. This will be manifested in a rich transmission profile for both positive and negative transverse momentum and the relative position of Fano resonances for different tilt strengths along the longitudinal and transverse directions. Due to the complexity of the situation, we leave this situation to future studies and instead note that when the transport can be constrained orthogonal to the tilt direction, a particularly simple relationship, as derived in this work, occurs.
Still, we expect that for insignificant longitudinal tilt strength as compared to the transverse component, our results will be qualitatively unaltered and the relation $\Delta(k_y,\tau_y) \to -\Delta(-k_y,\tau_y) $ continues to hold at least approximately. }

\textcolor{black}{Finally, in this section we briefly discuss a few possible forming mechanisms for the tilt term in the dispersion. A LBT is generically observed in monolayer graphene around the Dirac point. A tilted Dirac dispersion has been predicted to appear in various derivatives of monolayer graphene such as, quinoid-type graphene \cite{Goerbig08}, hydrogenated graphene \cite{Lu16},
8-Pmmn borophene \cite{Zabolotskiy16}, planar arrays of carbon nanotubes \cite{Polozkov19}, and artificial graphene \cite{mann2018manipulating}.  Uniaxial strain has also been reported to be useful to introduce a tilt in the energy dispersion of graphene \cite{Montambaux19}. Indirect evidence of a tilted LBT dispersion has also been found in organic semiconductors
\cite{katayama2006pressure,hirata2016observation}. Moreover,  tilted Dirac cones have also been  experimentally observed in photonic Lieb-kagome lattices \cite{Lang23}. For the case of QBT, as occuring in bilayer graphene, a tilt term can be engineered similarly as in monolayer graphene. Given the above discussion, it is evident that a tilt in LBT and QBT dispersions is not only a theoretical idea but also an experimentally feasible phenomena. This further makes our findings relevant from an experimental point of view. }

\section{Limitations}
\label{sec_new}

\textcolor{black}{Our study on the electronic transport through an oscillating potential well for systems hosting tilted LBT and QBT dispersions is useful for mesoscopic devices. The potential region can then be thought of as the central system, while the free regions on the left and right serve the role of leads.  However, in our analysis we consider a few assumptions in order to make our work, and specially the results, analytically and numerically tractable, which introduce some limitations as far as realistic mesoscopic devices are concerned. We discuss these below. }

\textcolor{black}{To begin with,  we consider an idealized situation where the potential well is infinitely extended in the $y$-direction  perpendicular to the transport direction. This assumption allows us to neglect  edge effects and the  single-particle wave-function can be considered as a plane wave in the $y$-direction without loss of generality. In reality, the central system would in a device have a finite width, which would introduce nodes in the wave-function at the edge. Therefore, one can expect a certain localization of wave-function with a non-plane wave-like solution in the transverse direction. For example, a sinusoidal nature of wave-function along transverse $y$-direction can be anticipated where $k_y$ is given by the width of the potential along $y$-direction. As a result, the variable $k_y$ no longer remains continuous, but rather there exist discrete allowed values of $k_y$. This causes the 
energies of the quasi-bound state to be a discrete function of $k_y$ as compared to the situation we analyze and hence the bound state energy gets modified in presence of 
a finite potential barrier along the transverse direction. 
As a consequence, the Fano resonance energy is expected to shift if the potential is not infinitely extended along the $y$-direction. However, the qualitative features remain unaltered, and especially in longer devices we expect a limited impact of the edges.}

\textcolor{black}{Moreover, we have been discussing the ideal situation so far where no dephasing phenomena takes place.  The role of the potential is to modify the nature of the wave-function inside the potential well only, while there is no other interactions present anywhere.  As a result, the single-particle wave-function always remains in a pure state, resulting in a sharp nature of the Fano resonance profiles.
However, in the context of transmission through a potential well as experienced in a realistic mesoscopic device some dephasing or decoherence phenomena is unavoidable.
To capture this latter situation, we believe one has to start with a density matrix formalism to probe the effect of decoherence in this time-dependent problem. One has to consider an interaction between the potential well and the underlying model, acting like the leads, such that the initial pure state evolves into a mixed state.  Once these interactions take place, the effect of decoherence comes into this dynamical setup, leading to a broadening of the Fano resonance peak-dip profile. In an effective manner, one can alternatively consider complex a non-Bloch form of the wave vector $k_x$ to handle such open quantum system exposed to the environment \cite{Kawabata19}.  This will  essentially modify the quasi-bound state energy  inside the potential well, again yielding a broadening of the Fano resonance profile \cite{Andreas10}. 
However, due to the significant addition of complexity, a study of dephasing/decoherence requires a separate investigation, which is beyond the scope of the present manuscript. Moreover, we emphasize that our results modeling the ideal situation creates an efficent and clear-cut result on which more realistic effects will act by producing energy shifts or broadening.}

%%%%%%%%%%%%%%%%%%%%%%%%%%%%%%%%%%%%%%%%%%%%%%%%%%%%%%%%%%%%%%%%%%%%
\section{Conclusions and outlook}
\label{sec5}
%%%%%%%%%%%%%%%%%%%%%%%%%%%%%%%%%%%%%%%%%%%%%%%%%%%%%%%%%%%%%%%%%%%%

Motivated by intriguing tilt mediated transport characteristics, including Hall responses, in solid state systems, we examine the transmission spectra, mimicking the effective current profile, and shot noise spectra for realistic low-energy dispersions through a quantum potnetial well.
To be precise, we scrutinize the effect of linear transverse tilt for LBT and QBT Hamiltonians when the quantum potential well, oriented longitudinally, is harmonically driven (Fig. \ref{fig:setup}).
We employ Floquet theory combined with a scattering matrix framework to study the emergence of Fano resonances due to Floquet sidebands that reflect multi-photon processes. Our study finds that the Fano resonance energy, caused by an overlap between the Floquet sidebands and quasi-bound state within the potential well,  shifts toward  lower (higher) values of energy when the magnitude of transverse  momentum increases for LBT (QBT) system, when keeping the tilt strength unaltered (Figs.~\ref{fig:graphene-transmission-posky}, \ref{fig:graphene-transmission-negky}, \ref{fig:qbt-transmission-posky}, and \ref{fig:qbt-transmission-negky}). On the other hand, for a given value of positive transverse momentum, positive (negative) tilt increases (decreases) the Fano resonance energy, irrespective of the underlying band dispersion of the system. Moreover, we show that the energies of the quasi-bound states are coupled to the tilt strength, and at the same time, some information about the underlying dispersion is also encoded in the energies of the quasi-bound state. Putting all these together, we show that a so-called tilt function, constructed out of the Fano resonance energy for two opposite values of transverse momenta, can clearly identify the tilt strength for the linear tilt without any a priori information.  
Furthermore, the qualitative differences between the Fano resonances can be used to distinguish between the LBT and QBT dispersion as the characteristic dip-peak structure becomes sharper and remains unaltered for LBT and QBT, respectively, when increasing the magnitude of transverse momentum.
Moreover, we extend our analysis to shot noise measurement to exploit the nature of the Fano resonance due to Floquet sidebands  more clearly (Figs. \ref{fig:graphene-shotnoise-posky}, \ref{fig:graphene-shotnoise-negky}, \ref{fig:qbt-shotnoise-posky}, and  \ref{fig:qbt-shotnoise-negky}). 
The differential shot noise identifies the Fano resonance energy clearly as it captures the contribution from the dominant Floquet sidebands. Importantly, the differential shot noise and transmission spectra both exhibit qualitatively similar behavior. 

Our work yields a qualitative impression of the transmission and shot noise spectra for materials with  LBT and QBT low-energy dispersions around the Fermi level, in presence of a harmonically driven chemical potential that represents the quantum well locally at each site.  From the experimental side, shot noise  spectra are investigated also in the ultracold atoms \cite{Gajdacz16}, as well as solid state systems \cite{tikhonov2015shot}.
We believe that our theoretical analysis, based on  microscopic mechanisms, will be useful to understand the experimental findings for periodically driven systems in general.
In addition, in the future, variation of the tilt parameter could be instrumental to engineer 
Fano resonance energy and the shape of the corresponding peak-dip structure. Furthermore, Floquet scattering properties can also be investigated in the presence of disorder \cite{Okugawa22},  magnetic fields \cite{Saha21,Nag21flat}, and electric fields.

\section{Acknowledgement} 

TN would like to thank Banasri Basu, Arnab Maity, Souvik Das, Rajib  Sarkar, and Anirudha Menon for useful discussions. AG thanks Sandip Bera for discussions on shot noise. TN thanks Ipsita Mondal for the very initial discussions on the Floquet scattering techniques.  This work was financially supported by  the Swedish Research Council (Vetenskapsr\aa det Grant No.~2018-03488), the Knut and Alice Wallenberg Foundation through the Wallenberg Academy Fellows program and the project grant KAW 2019.0068.

%%%%%%%%%%%%%%%%%%%%%%%%%%%%%%%%%%%%%%%%%%%%%%%%%%%%%%%%%%%%%%%%%%%%
\appendix
%%%%%%%%%%%%%%%%%%%%%%%%%%%%%%%%%%%%%%%%%%%%%%%%%%%%%%%%%%%%%%%%%%%%
%\begin{widetext}

\section{S-matrix for LBT}\label{appA}
In this Appendix we show the derivation for the ${\bm S}$-matrix for the LBT case as discussed in Sec.~\ref{sec2_1}. We start by expressing the continuity 
conditions for the wave functions given in Eq. (\ref{eq:tiltgraphenewf}) at the boundaries \(x = \pm L/2\):
\begin{align}
    \bm{A}_n^i&e^{-ik_{xn}L/2}+\bm{A}_n^oe^{ik_{xn}L/2} = \nonumber \\ 
    &\sum_{m=-\infty}^{\infty}\bqty{a_me^{-iq_mL/2}+b_me^{iq_mL/2}}J_{n-m},
\end{align}
\begin{align}
    &\bm{A}_n^is_ne^{-ik_{xn}L/2+i\phi_n}-\bm{A}_n^os_ne^{ik_{xn}L/2-i\phi_n} =\nonumber\\
    & \sum_{m=-\infty}^{\infty}\bqty{s'_ma_me^{-iq_mL/2+i\theta_m}-s'_mb_me^{iq_mL/2-i\theta_m}}J_{n-m},
\end{align}
\begin{align}
    \bm{B}_n^i&e^{-ik_{xn}L/2}+\bm{B}_n^oe^{ik_{xn}L/2} = \nonumber\\
    &\sum_{m=-\infty}^{\infty}\bqty{a_me^{iq_mL/2}+b_me^{-iq_mL/2}}J_{n-m},
\end{align}
and
\begin{align}
    &-\bm{B}_n^is_ne^{-ik_{xn}L/2-i\phi_n}+\bm{B}_n^os_ne^{ik_{xn}L/2+i\phi_n} = \nonumber\\
    &\sum_{m=-\infty}^{\infty}\bqty{s'_ma_me^{iq_mL/2+i\theta_m}-s'_mb_me^{-iq_mL/2-i\theta_m}}J_{n-m},
\end{align}
respectively. For positive incident energy above a threshold value, we can find that $s_n$ always remains positive. This can lead to a simplification in the continuity condition. However, we proceed with the general situation where no conditions on the range of incident energy is imposed. Defining 
\begin{align}
    \pqty{\bm{M}_{sa}^\pm}_{nm} =& [e^{-iq_mL/2}(s_ne^{-i\phi_n}+s_m'e^{i\theta_m})\nonumber\\
    &\pm e^{iq_mL/2}(s_ne^{i\phi_n}-s_m'e^{i\theta_m})]J_{n-m},\nonumber\\
    \pqty{\bm{M}_{sb}^\pm}_{nm} =& [e^{iq_mL/2}(s_ne^{-i\phi_n}-s_m'e^{-i\theta_m})\nonumber\\
    &\pm e^{-iq_mL/2}(s_ne^{i\phi_n}+s_m'e^{-i\theta_m})]J_{n-m},\nonumber\\
    \pqty{\bm{M}_r}_{nm} =& 2\cos(\phi_n)s_ne^{-ik_{xn}L/2}\delta_{nm},\nonumber\\
    \pqty{\bm{M}_i}_{nm} =& e^{-ik_{xn}L}\delta_{nm},\\
    \pqty{\bm{M}_c^\pm}_{nm} =& e^{-i(k_{xn}\pm q_m)L/2}J_{n-m},\nonumber
\end{align}
as well as 
\begin{align}
    \bm{a}_A =& \bqty{\pqty{\bm{M}_{sb}^+}^{-1}\bm{M}_{sa}^+-\pqty{\bm{M}_{sb}^-}^{-1}\bm{M}_{sa}^-}^{-1}\nonumber\\
    &\times\bqty{\pqty{\bm{M}_{sb}^+}^{-1}-\pqty{\bm{M}_{sb}^-}^{-1}}\bm{M}_r,\nonumber\\
    \bm{a}_B =& \bqty{\pqty{\bm{M}_{sb}^+}^{-1}\bm{M}_{sa}^+-\pqty{\bm{M}_{sb}^-}^{-1}\bm{M}_{sa}^-}^{-1}\nonumber\\
    &\times\bqty{\pqty{\bm{M}_{sb}^+}^{-1}+\pqty{\bm{M}_{sb}^-}^{-1}}\bm{M}_r,\\
    \bm{b}_A =& \bqty{\pqty{\bm{M}_{sa}^+}^{-1}\bm{M}_{sb}^+-\pqty{\bm{M}_{sa}^-}^{-1}\bm{M}_{sb}^-}^{-1}\nonumber\\
    &\times\bqty{\pqty{\bm{M}_{sa}^+}^{-1}-\pqty{\bm{M}_{sa}^-}^{-1}}\bm{M}_r,\nonumber\\
    \bm{b}_B =& \bqty{\pqty{\bm{M}_{sa}^+}^{-1}\bm{M}_{sb}^+-\pqty{\bm{M}_{sa}^-}^{-1}\bm{M}_{sb}^-}^{-1}\nonumber\\
    &\times\bqty{\pqty{\bm{M}_{sa}^+}^{-1}+\pqty{\bm{M}_{sa}^-}^{-1}}\bm{M}_r,\nonumber
\end{align}
one can show that 
\begin{align}
        \begin{cases}
        \bm{A}^o =& \pqty{\bm{M}_c^+\bm{a}_A+\bm{M}_c^-\bm{b}_A-\bm{M}_i}\bm{A}^i\\
        &+\pqty{\bm{M}_c^+\bm{a}_B+\bm{M}_c^-\bm{b}_B}\bm{B}^i,\\
        \bm{B}^o =& \pqty{\bm{M}_c^-\bm{a}_A+\bm{M}_c^+\bm{b}_A}\bm{A}^i\\
        &+\pqty{\bm{M}_c^-\bm{a}_B+\bm{M}_c^-\bm{b}_B-\bm{M}_i}\bm{B}^i.
    \end{cases}
\end{align}
The ${\bm S}$-matrix is then defined as 
\begin{align}
\bm{S} = \pmqty{\bm{M_{AA}} & \bm{M_{AB}} \\ \bm{M_{BA}} & \bm{M_{BB}}}
\end{align}
with $\bm{M_{AA}}=\bm{M}_c^+\bm{a}_A+\bm{M}_c^-\bm{b}_A-\bm{M}_i$, $\bm{M_{AB}}=\bm{M}_c^+\bm{a}_B+\bm{M}_c^-\bm{b}_B$, $\bm{M_{BA}}=\bm{M}_c^-\bm{a}_A+\bm{M}_c^+\bm{b}_A$, and $\bm{M_{BB}}=\bm{M}_c^-\bm{a}_B+\bm{M}_c^-\bm{b}_B-\bm{M}_i$. This results in 
\begin{align}
    \pmqty{\bm{A}^o\\\bm{B}^o} =  \bm{S}\pmqty{\bm{A}^i \\ \bm{B}^i},
\end{align}
which is equivalent to Eq.~\eqref{eq:S-matrix-def} in the main text for the LBT dispersion when we contract the summation over Floquet side bands.
To calculate the total transmission in Eq. (\ref{eq:transmission-final}), we numerically compute $\bm{M_{BA}} \equiv s_{RL} \equiv t$ (Eq.~(\ref{eq:S-matrix-comps})). In order to evaluate the shot noise, given in Eq.~(\ref{eq:shotnoise-final}), we calculate $\bm{M_{AA}}\equiv s_{LL} \equiv r$, $\bm{M_{AB}} \equiv s_{LR} \equiv t'$, $\bm{M_{BA}} \equiv s_{RL} \equiv t$, and $\bm{M_{BB}} \equiv s_{RR} \equiv r'$ numerically. 

%%%%%%%%%%%%%%%%%%%%%%%%
\section{S-matrix for QBT }\label{appB}
%%%%%%%%%%%%%%%%%%%%%%%%%
In this Appendix we show the derivation for the ${\bm S}$-matrix for the QBT case as discussed in Sec.~\ref{sec2_2}.
We start by expressing the continuity 
conditions for the wave functions given in Eq. (\ref{eq:tiltwf})  at the boundaries \(x = \pm L/2\):
\begin{align}
        \eta_{1,n}&\pqty{\bm{A}^i_ne^{-ik_{xn}L/2}+\bm{A}^o_ne^{ik_{xn}L/2}} = \nonumber\\
        &\sum_{m=-\infty}^\infty[\eta_{2,m}\left(a_me^{-iq_mL/2}+b_me^{iq_mL/2}\right)\Theta_m^+\nonumber\\
        &+\eta_{3,m}\left(a_me^{-iq_mL/2}+b_me^{iq_mL/2}\right)\Theta_m^-]J_{n-m}\pmqty{\frac{V_1}{\hbar\omega}},
\end{align}
\begin{align}
        \eta_{1,n}&\frac{k_{xn}}{k_y}\pqty{\bm{A}^i_ne^{-ik_{xn}L/2}-\bm{A}^o_ne^{ik_{xn}L/2}} = \nonumber\\
        &\sum_{m=-\infty}^\infty [\eta_{2,m}\frac{q_m}{k_y}(a_me^{-iq_mL/2}-b_me^{iq_mL/2})\Theta_m^+\nonumber\\
        &+\eta_{3,m}\frac{k_y}{q_m}\left(-a_me^{-iq_mL/2}+b_me^{iq_mL/2}\right)\Theta_m^-]J_{n-m}\pmqty{\frac{V_1}{\hbar\omega}},
\end{align}
\begin{align}
    \eta_{1,n}&\pqty{\bm{B}^o_ne^{ik_{xn}L/2}+\bm{B}^i_ne^{-ik_{xn}L/2}} = \nonumber\\
    &\sum_{m=-\infty}^\infty[\eta_{2,m}\left(a_me^{iq_mL/2}+b_me^{-iq_mL/2}\right)\Theta_m^+\nonumber\\
    &+\eta_{3,m}\left(a_me^{iq_mL/2}+b_me^{-iq_mL/2}\right)\Theta_m^-]J_{n-m}\pmqty{\frac{V_1}{\hbar\omega}},
\end{align}
and 
\begin{align}
    \eta_{1,n}&\frac{k_{xn}}{k_y}\pqty{\bm{B}^o_ne^{ik_{xn}L/2}-\bm{B}^i_ne^{-ik_{xn}L/2}} = \nonumber\\
    &\sum_{m=-\infty}^\infty [\eta_{2,m}\frac{q_m}{k_y}\left(a_me^{iq_mL/2}-b_me^{-iq_mL/2}\right)\Theta_m^+\nonumber\\
    &+\eta_{3,m}\frac{k_y}{q_m}\left(-a_me^{iq_mL/2}+b_me^{-iq_mL/2}\right)\Theta_m^-]J_{n-m}\pmqty{\frac{V_1}{\hbar\omega}},
\end{align}
respectively. Note that  $\Theta^+_n$ ($\Theta^-_n$)  reduces to unity (zero) in the region $x> |L/2|$ for the incident energy beyond a threshold value. This causes the simplified continuity conditions as demonstrated above.  Defining 
\begin{align}
    (M_r)_{nm} =& 2\eta_{1,n}e^{-ik_{xn}L/2} \delta_{nm}, \nonumber\\
    (M_s^\pm)_{nm} =& J_{n-m}[\eta_{2,m}\left((1+\frac{q_m}{k_{xn}})e^{-iq_mL/2}\pm(1-\frac{q_m}{k_{xn}})e^{iq_mL/2}\right)\Theta_m^+ \nonumber\\
    +\eta_{3,m}&\left((1-\frac{k_y^2}{k_{xn}q_m})e^{-iq_mL/2}\pm (1+\frac{k_y^2}{k_{xn}q_m})e^{iq_mL/2}\right)\Theta_m^-], \nonumber\\
    (M_c^\pm)_{nm} =& \frac{J_{n-m}}{\eta_{1,n}}(\eta_{2,m}\Theta_m^++\eta_{3,m}\Theta_m^-)e^{-i(k_{xn}\pm q_m)L/2}, \nonumber\\
    (M_i)_{nm} =& e^{-ik_{xn}L}\delta_{nm},
\end{align}
and 
\begin{align*}
    \bm{a_A} &= \frac{1}{2}((\bm{M}_s^+)^{-1}+(\bm{M}_s^-)^{-1})\cdot \bm{M}_r,\\
    \bm{a_B} &= \frac{1}{2}((\bm{M}_s^+)^{-1}-(\bm{M}_s^-)^{-1})\cdot \bm{M}_r,\\
    \bm{b_A} &= \bm{a_B},\quad    \bm{b_B} = \bm{a_A},
\end{align*}
one can show that
\begin{align}
\begin{cases}
\bm{A}^o =& (\bm{M}_c^+\cdot\bm{a_A}+\bm{M}_c^-\cdot\bm{b_A}-\bm{M}_i)\cdot\bm{A}^i\\
    &+ (\bm{M}_c^+\bm{a_B}+\bm{M}_c^-\bm{b_B}) \cdot \bm{B}^i\\
    \bm{B}^o =& (\bm{M}_c^-\cdot\bm{a_A}+\bm{M}_c^+\cdot\bm{b_A})\cdot \bm{A}^i \\
    &+(\bm{M}_c^-\cdot\bm{a_B}+\bm{M}_c^+ \cdot \bm{b_B}-\bm{M}_i) \cdot \bm{B}^i.
\end{cases}
\end{align}
The ${\bm S}$-matrix is then defined as 
\begin{align}
\bm{S} = \pmqty{\bm{M_{AA}} & \bm{M_{AB}} \\ \bm{M_{BA}} & \bm{M_{BB}}}
\end{align}
such that 
\begin{align}
    \pmqty{\bm{A}^o\\\bm{B}^o} =  \bm{S}\pmqty{\bm{A}^i \\ \bm{B}^i}.
\end{align}
The explicit expression for the $\bm{M}$ matrices in the this case remains unaltered as compared to the LBT cases. 
Therefore, similar to the previous case of LBT, we numerically compute $\bm{M_{AA}}\equiv s_{LL} \equiv r$, $\bm{M_{AB}} \equiv s_{LR} \equiv t'$, $\bm{M_{BA}} \equiv s_{RL} \equiv t$, and $\bm{M_{BB}} \equiv s_{RR} \equiv r'$ to evaluate the transmission and shot noise spectra. 

%\end{widetext}

%%%%%%%%%%%%%%%%%%%%%%%%%%%%%%%%%%%%%%%%%%%%%%%%%%%%%%%%%%%%%%%%%%%%
\bibliography{ref}
%%%%%%%%%%%%%%%%%%%%%%%%%%%%%%%%%%%%%%%%%%%%%%%%%%%%%%%%%%%%%%%%%%%%

\end{document}